\documentclass{amsart}
\usepackage{amssymb}
\usepackage{amsfonts}
\usepackage{graphicx}
\usepackage{epstopdf}

\newcommand{\beq}{\begin{equation}}
\newcommand{\eeq}{\end{equation}}

\begin{document}

\title{Universality in Chern-Simons theory}
       
   \author{R. L. Mkrtchyan}
       \address{Yerevan Physics Institute, 2 Alikhanian Brothers St., Yerevan, 0036, Armenia}
           \email{mrl@web.am}
           
        \author{A. P. Veselov}
       \address{Department of Mathematics,
        Loughborough University, Loughborough,
        Leicestershire, LE11 3TU, UK and
       Moscow State University, Russia}
       \email{A.P.Veselov@lboro.ac.uk}

\maketitle




{\small  {\bf Abstract.} We show that the perturbative part of the partition function in the Chern-Simons theory on a 3-sphere as well as the central charge and expectation value of the unknotted Wilson loop in the adjoint representation can be expressed in terms of the universal Vogel's parameters $\alpha, \beta, \gamma.$ The derivation is based on certain  generalisations of the Freudenthal-de Vries strange formula. }

\section{Introduction}

We are going to discuss the universality phenomenon for the gauge theories in the example of the Chern-Simons theory, using the notion of {\it universal Lie algebra} introduced by Vogel \cite{V0,V}. Mathematically the latter is a certain tensor category having Vogel plane as a moduli space with special points corresponding to all simple Lie algebras (see below). 

One can think on a universal gauge theory as the one based on the universal Lie algebra rather than on a concrete simple Lie algebra. It is not clear how to make precise meaning of this, but what certainly makes sense is to talk about universal formulae in the theory. Namely, we say that a quantity in the gauge theory is {\it universal} if it can be expressed as an analytic (say, rational) function of Vogel's parameters. In particular, the universal Chern-Simons theory (UCS) is dealing with the universal quantities within this theory.  

The elements of the universality were known before Vogel. In physics it goes back at least to 't Hooft  \cite{H1}, who extended the $SU(N)$ gauge theory to arbitrary $N.$ Witten's analytic continuation of Chern-Simons theory \cite{W2} is one of the most recent important developments in this direction. Part of the universality is $n \rightarrow -n$ duality between $SO(2n)$ and $Sp(2n)$ gauge theories \cite{Ki, Mkr, Cvitbook, VM}.

Another source of universality came from the attempts to understand better
the classification of the simple Lie algebras, see \cite{Cvitbook, Ho, An}.
In particular, the work by  Deligne et al. \cite{Del,DM} on the exceptional series is closely related to Vogel's approach, which was motivated by the knot theory. A close relation of knot invariants with Chern-Simons theory became clear after the work of Witten \cite{W1} (see e.g. Bar-Natan \cite{BarNatan1, BarNatan2}). A mathematically rigorous alternative to Chern-Simons theory as a theory of invariants of 3-manifolds based on quantum groups was proposed by Reshetikhin and Turaev \cite{RT,T}.

Recall that {\it Vogel plane} is the quotient space $P^2/S_3$ of projective plane with projective coordinates $\alpha, \beta$ and $\gamma$ (Vogel's parameters) by the symmetric group $S_3$ acting by permutations of the parameters. The projective nature of the parameters corresponds to the choice of the invariant bilinear form on simple Lie algebra, which is known to be unique up to a multiple. The parameters corresponding to simple Lie algebras are given in Table 1, where we have chosen the normalisation with $\alpha=-2.$ This normalisation corresponds to the so-called {\it minimal bilinear form}, when the square of the length of the maximal root is chosen to be 2 and  the value of the corresponding quadratic Casimir operator on the adjoint representation $$C_2=2t=2h^\vee,$$ where $h^\vee$ is the {\it dual Coxeter number} and $t=\alpha+\beta+\gamma.$ The canonical Cartan-Killing form corresponds to $C_2=1$, so $t=1/2.$

Vogel's parameters have the following meaning \cite{V0}: take symmetric part of the tensor square of the adjoint representation of $\mathfrak g$, decompose it into three (two for the exceptional Lie algebras) irreducible representations, then the eigenvalues of the second Casimir on them are $4t-2\alpha, 4t-2\beta, 4t-2\gamma.$

A remarkable fact is that many numerical characteristics of simple Lie algebras can be expressed in terms of only these 3 parameters by the formulae, which we will call universal.
 An example is the dimension of $\mathfrak g$ given by Vogel's formula
\beq
\label{f3}
dim \, \mathfrak {g} = \frac{(\alpha-2t)(\beta-2t)(\gamma-2t)}{\alpha\beta\gamma}.
\eeq
(see \cite{LM1,LM2} for further developments in this direction). Another class of universal quantities are the eigenvalues of certain natural Casimir operators in adjoint representation of  $\mathfrak g$ (see \cite{MSV}). 

\begin{table}[h]  
\caption{Vogel's parameters for simple Lie algebras}     
\begin{tabular}{|c|c|c|c|c|c|}
\hline
Type & Lie algebra  & $\alpha$ & $\beta$ & $\gamma$  & $t=h^\vee$\\   
\hline    
$A_n$ &  $\mathfrak {sl}_{n+1}$     & $-2$ & 2 & $(n+1) $ & $n+1$\\
$B_n$ &   $\mathfrak {so}_{2n+1}$    & $-2$ & 4& $2n-3 $ & $2n-1$\\
$C_n$ & $ \mathfrak {sp}_{2n}$    & $-2$ & 1 & $n+2 $ & $n+1$\\
$D_n$ &   $\mathfrak {so}_{2n}$    & $-2$ & 4 & $2n-4$ & $2n-2$\\
$G_2$ &  $\mathfrak {g}_{2}  $    & $-2$ & $10/3 $& $8/3$ & $4$ \\
$F_4$ & $\mathfrak {f}_{4}  $    & $-2$ & $ 5$& $ 6$ & $9$\\
$E_6$ &  $\mathfrak {e}_{6}  $    & $-2$ & $ 6$& $ 8$ & $12$\\
$E_7$ & $\mathfrak {e}_{7}  $    & $-2$ & $ 8$& $ 12$ & $18$ \\
$E_8$ & $\mathfrak {e}_{8}  $    & $-2$ & $ 12$& $20$ & $30$\\
\hline  
\end{tabular}
\end{table}

Consider now the Chern-Simons gauge theory with the action
\begin{equation}\label{CSA}
S(A)=\frac{\kappa}{4 \pi} \int_M  Tr \left( A \wedge d A + \frac{2}{3} A\wedge A \wedge A \right).
\end{equation}
Here $A$ is $\mathfrak g$-valued 1-form on a 3-dimensional manifold $M$ (which we choose here to be $S^3$) and $Tr$ denotes some invariant bilinear form on a simple Lie algebra $\mathfrak g$ (see \cite{W1}). We will not fix the choice of such a form allowing coupling constant $\kappa$ to change accordingly under replacing the form to preserve the action. This means that the universal Chern-Simons theory depends on 4 parameters $\alpha, \beta, \gamma, \kappa$ defined up to a common multiple, where $\alpha,\beta, \gamma$ are Vogel's parameters. In fact it is more convenient to replace $\kappa$ by 
$$\delta=\kappa+t=\kappa+\alpha+\beta+ \gamma.$$
Note that we have permutation symmetry in the first 3 parameters, so the corresponding moduli space is
a quotient of the 3D projective space $P^3/S_3.$

An example of the universality in Chern-Simons theory  is the following formula for the central charge $c,$  which governs the phase change of partition function under the change of trivialisation of tangent bundle of $M$: 
\beq \label{cg}
c=\frac{\kappa(\alpha-2t)(\beta-2t)(\gamma-2t)}{\alpha\beta\gamma(\kappa+\alpha+\beta+ \gamma)}=\frac{(\delta-t)(\alpha-2t)(\beta-2t)(\gamma-2t)}{\alpha\beta\gamma\delta}.
\eeq
Indeed, it is known that if we choose minimal bilinear form, so that $\kappa=k$ (which needs to be integer in that case) and
$t=h^\vee$ is the dual Coxeter number, then the central charge is given by (see \cite{W1})
\beq \label{cg1}
c=k\frac{ dim(g)}{k+h^\vee}.
\eeq	
Taking into account Vogel's formula (\ref{f3}) we can rewrite this in universal parameters as (\ref{cg}).

We are going to show that the same is true for the perturbative part of the Chern-Simons partition function and for the expectation value of unknotted Wilson loop in $S^3$ in the adjoint representation. Mathematically this is based on the existence of the universal formulae for the sums 
$$p_{2m}(\mathfrak g)=\sum_{\mu \in R_+} (\mu, \rho)^{2m},$$
where $R_+$ is the positive part of the root system of the Lie algebra and $\rho$ is the half-sum of positive roots (Weyl's vector). We are using the unusual notations for roots since $\alpha$ was taken by Vogel.
The first universal formula has the form
\beq \label{2}
\sum_{\mu \in R_+} (\mu, \rho)^{2}= \frac{t^2}{12} dim \, \mathfrak {g}, 
\eeq
which is a homogeneous form of the {\it Freudenthal-de Vries strange formula}
$$ <\rho, \rho>=\frac{1}{24} dim \, \mathfrak {g},$$ where $< , >$ is the Cartan-Killing form (see \cite{FdV}).
We show that similar universal formulae exist for all $p_{2m}(\mathfrak g)$, in particular
\beq \label{4}
\sum_{\mu \in R_+} (\mu, \rho)^{4}= \frac{t(18t^3-3t t_2 +t_3)}{480} dim \, \mathfrak {g}, 
\eeq
where $t_2=\alpha^2+\beta^2+\gamma^2,\,\, t_3=\alpha^3+\beta^3+\gamma^3.$

Our results can be interpreted in terms of the corresponding (nontwisted) affine Lie algebras \cite{Kac},
so one can view this work also as an extension of the Vogel map to the infinite-dimensional case.

\section{Chern-Simons partition function of $S^3$}

The Chern-Simons partition function 
$$Z(M)=\int DA exp\left(\frac{ik}{4\pi}\int_M Tr \left( A \wedge d A + \frac{2}{3} A\wedge A \wedge A \right)\right)$$
in the case of sphere $M=S^3$ is known explicitly \cite{W1}. 
Let $\mathfrak h$ be Cartan subalgebra of the Lie algebra $\mathfrak g$, $r$ be its rank, $Q, P \subset \mathfrak h^*$ be the root and weight lattices respectively, $Q^\vee \subset \mathfrak h$ be the coroot lattice and choose the minimal invariant bilinear form on $\mathfrak g$. In that case we have (see e.g. \cite{DiF}, (14.241)):
\beq
\label{zz}
Z=Z(S^3) = Vol(Q^\vee)^{-1}(k+h^\vee)^{-r/2}\prod_{\mu \in R_+}
2\sin \frac{\pi (\mu,\rho)}{( k+h^\vee)}.
\eeq
where $Vol(Q^\vee)$ means the volume of the fundamental domain of $Q^\vee$ with respect to the minimal invariant form.
An arbitrary form corresponds to replacing $k+h^\vee$ by $\delta$ with the corresponding rescaling of $Vol(Q^\vee)$:
\beq
\label{zz*}
Z=Z(S^3) = Vol(Q^\vee)^{-1}\delta^{-r/2}\prod_{\mu \in R_+}
2\sin \frac{\pi (\mu,\rho)}{\delta}.
\eeq
Let's rewrite it as the product $Z=Z_{1} Z_2,$ where
\beq
\label{zz1}
Z_1=Vol(Q^\vee)^{-1}\delta^{-r/2}\prod_{\mu \in R_+}
\frac{2\pi (\mu,\rho)}{\delta}
\eeq
and
\beq
\label{zz2}
Z_2=\prod_{\mu \in R_+}
\sin \frac{\pi (\mu,\rho)}{\delta}/\frac{\pi (\mu,\rho)}{\delta}.
\eeq

The first factor $Z_1$ has a clear geometric meaning (cf. \cite{W1, OV}):
\beq
\label{zz12}
Z_1=\frac{(2\pi  \delta^{-1/2})^{dim \, \mathfrak g}} {Vol(G)},
\eeq
where $Vol(G)$ is the volume of the corresponding compact simply connected group $G$ (which is $SU(n)$ and $Sp(n)$ in the unitary and symplectic cases and the double cover $Spin(n)$ of $SO(n)$ in the orthogonal case).
Indeed, 
$$
Z_1=Vol(Q^\vee)^{-1}\delta^{-r/2}\prod_{\mu \in R_+}
\frac{2\pi (\mu,\rho)}{\delta}=\frac{\pi^N}{Vol(Q^\vee)\delta^{\frac{dim \, \mathfrak g}{2}}}\prod_{\mu \in R_+}
\frac{2(\mu,\rho)}{(\mu,\mu)} \prod_{\mu \in R_+} \frac{4}{(\mu^\vee,\mu^\vee)},
$$
where $\mu^\vee=\frac{2\mu}{(\mu,\mu)}$, $N$ is the number of positive roots, so that $r+2N= dim \, \mathfrak g.$ Now we can use Macdonald's formula \cite{Ma, Hash} for the volume of the group $G$
\beq
\label{mac}
Vol(G)=Vol(Q^\vee)\prod_{i=1}^r \frac{2\pi^{m_i+1}}{m_i!} \prod_{\mu \in R_+}(\mu^\vee,\mu^\vee),
\eeq
where $m_i$ are the exponents of the Lie algebra $\mathfrak g$ (which are one less than the degrees of the basic Casimir elements), satisfying the identity (see \cite{Bour})
$$\sum_{i=1}^r m_i=N.$$
To come to (\ref{zz12}) we need one more important  identity due to Steinberg
\beq
\label{hash}
\prod_{\mu \in R_+}\frac{2(\mu,\rho)}{(\mu,\mu)} = \prod_{i=1}^r m_i!
\eeq
(see Appendix to \cite{H}).

Let us look now at the second factor $Z_2$, which can be called the perturbative part of the partition function (see \cite{OV}).  Consider the corresponding free energy $F_2=- \ln Z_2$. 
Using infinite product representation
$$
\sin \pi x= \pi x \prod^{\infty}_{n=1} (1-(\frac{x}{n})^2)
$$
we have
$$ \ln \frac {\sin \pi x}{\pi x}= \sum ^{\infty}_{n=1} \ln (1-(\frac{x}{n})^2)=\sum ^{\infty}_{n=1} \sum ^{\infty}_{m=1} \frac{1}{m}\frac{x^{2m}}{n^{2m}}=\sum ^{\infty}_{m=1} \frac{\zeta(2m)}{m} x^{2m},$$
where $\zeta(z)$ is the Riemann zeta-function. The values $\zeta(2m)$ for positive integer $m$ are known explicitly:
$$\zeta(2m)=(-1)^{m+1}\frac{B_{2m}(2\pi)^{2m}}{2(2m)!},$$
where $B_{2m}$ are the Bernoulli numbers.
Thus the perturbative part of free energy is
\beq
\label{free}
F_2=\sum ^{\infty}_{m=1} \frac{\zeta(2m)}{m} \sum_{\mu \in R_+} \left(\frac{(\mu,\rho)}{\delta}\right)^{2m}.
\eeq
To show its universality we should express the sums $\sum_{\mu \in R} (\mu,\rho)^{2m}$ in terms of Vogel's parameters.

It is more convenient for us to consider the sums over all roots
$$p_k=\sum_{\mu \in R} (\mu,\rho)^{k}.$$
We have $p_k=0$ for all odd $k$ and $p_{2m}=2 \sum_{\mu \in R_+} (\mu,\rho)^{2m}$ and $p_0=2N$ is the total number of roots. We drop this $p_0$ term and consider 
the following exponential generating function
$$F(x)=\sum_{k=1}^{\infty}\frac{p_{k}}{k!} x^{k}=
\sum_{\mu\in R} (e^{x(\mu,\rho)}-1).$$
We claim that it can be expressed in terms of the Vogel's parameters as follows
\beq
\label{gene}
F(x)=\frac{\sinh(x\frac{\alpha-2t}{4})}{\sinh(\frac{x\alpha}{4})}\frac{\sinh(x\frac{\beta-2t}{4})}{\sinh(x\frac{\beta}{4})}\frac{\sinh(x\frac{\gamma-2t}{4})}{\sinh(x\frac{\gamma}{4})}-\frac{(\alpha-2t)(\beta-2t)(\gamma-2t)}{\alpha\beta\gamma}
\eeq

The proof follows the idea from \cite{FdV, DiF}. Consider the value of character of the adjoint representation at the point $x\rho$ according to Weyl formula
\beq
\chi_\theta (x\rho)= \frac{D_{\theta+\rho}(x\rho)}{D_\rho(x\rho)}=\frac{D_\rho(x(\theta+\rho))}{D_\rho(x\rho)}=\prod _{\mu\in R_+}\frac{\sinh(x(\mu,\theta+\rho)/2)}{\sinh(x(\mu,\rho)/2)}
\eeq
where $\theta$ is the maximal root and
$$
D_\rho=\sum_{w\in W} \epsilon (w)e^{w\rho}=\prod _{\mu \in R_+} (e^{\alpha/2}-e^{-\alpha/2}).
$$
On the other hand this value is $r+\sum_{\mu\in R} e^{x(\mu,\rho)}$, so we have the equality
\beq \label{main}
r+\sum_{\mu\in R} e^{x(\mu,\rho)}=\prod _{\mu\in R_+}\frac{\sinh(x(\mu,\theta+\rho)/2)}{\sinh(x(\mu,\rho)/2)}.
\eeq
Now we need the following key property of the Vogel parameters:
\beq \label{key}
\prod _{\mu\in R_+}\frac{\phi((\mu,\theta+\rho))}{\phi((\mu,\rho))}=\frac{\phi((\alpha-2t)/2)}{\phi(\alpha/2)}\frac{\phi((\beta-2t)/2)}{\phi(\beta/2)}\frac{\phi((\gamma-2t)/2)}{\phi(\gamma/2)}
\eeq
for any even or odd function $\phi(x)$. The proof is based on the observation that the roots with non-zero scalar product with $\theta$ (the only one to contribute to the left hand side) can be
organized in three "strings" with the cancellations between consecutive numerators and denominators within each string, leaving only three of them in the right hand side 
(cf. \cite{LM1,MSV}).
Subtracting from both sides the Vogel's expression for the dimension gives (\ref{gene}), which implies universality of the perturbative part of free energy. 

Expanding both sides (\ref{gene}) in $x$ in the leading second order we have the Freudenthal-de Vries strange formula. The next two orders give (\ref{4}) and
\beq \label{6}
\sum_{\mu \in R_+} (\mu,\rho)^{6}=\frac{t(396t^5-157t^3t_2+15tt_2^2+39t^2 t_3 - 5 t_2 t_3)}{16128} dim \, \mathfrak g.
\eeq
 
 So what's about the universality of the full partition function ?
 As we have seen it is equivalent to the question of whether the volume of the corresponding group $G$ can be expressed in terms of the Vogel parameters or not. This volume is a product $$Vol(G)=Vol(T) Vol(G/T)$$ of the volumes of the maximal torus $T \subset G$ and of the corresponding flag variety $G/T.$
 The identity (\ref{hash}) and the universality of the $p_{2m}(\mathfrak g)$ suggest that at least for the volume $Vol(G/T)$ the answer should be positive, but it is most likely that we will have to go beyond the class of rational functions. We hope to come back to this very interesting question soon.

\section{Expectation value of Wilson loop}

Important gauge invariant quantity of gauge theories is the  expectation value of Wilson loop is
\beq \label{ZC}
<W(C)>=\frac{1}{Z} \int dA e^{iS(A)} W(C), 
\eeq
where
$$W(C)= Tr P(\exp{\int A_\mu dx^\mu})$$
with $A_\mu$ taken in some representation of $\mathfrak g$. 
For the unknotted curve $C$ on $S^3$ and representation with highest weight $\lambda$ the answer is known to be 
$$<W(C)>=\prod_{\mu\in R^+}
\frac{sin(\frac{\pi (\mu,\lambda+\rho)}{( k+h^\vee)}}{sin(\frac{\pi (\mu,\rho)}{( k+h^\vee)})},$$
with the minimal choice of the bilinear form (see \cite{W1,DiF}).
If we now choose the adjoint representation with $\lambda=\theta$ and free the form using the same arguments as before we come to the universal expression
\beq
\label{univer}
<W(C)>=\frac{sin(\frac{\pi (\alpha-2t)}{2\delta})}{sin(\frac{\pi \alpha}{2\delta})}\frac{sin(\frac{\pi (\beta-2t)}{2\delta})}{sin(\frac{\pi \beta}{2\delta})} \frac{sin(\frac{\pi (\gamma-2t)}{2\delta})}{sin(\frac{\pi \gamma}{2\delta})}.
\eeq
How to extend this to non-trivial knots and other representations is an interesting question.

We would like to note that this expression is invariant under the following involution 
\beq
\label{level-rank}
\sigma: (\alpha,\beta,\gamma,\delta)\rightarrow (-\alpha,-\beta,2\delta-\gamma,\delta).
\eeq
It resembles the {\it level-rank duality} (see  \cite{NS} and references therein) mixing level $k$ with parameters of group, although apparently different. 
Together with permutation symmetry $S_3$ of parameters $\alpha,\beta,\gamma,$ this involution generates an infinite semi-direct product group $\mathfrak K=S_3 \ltimes(\mathbb Z_2 \ltimes \mathbb Z^2).$ Its normal subgroup $\mathbb Z_2 \ltimes \mathbb Z^2$ consists of the shear transformations
$$(\alpha,\beta,\gamma,\delta)\rightarrow (\alpha+2a\delta,\beta+2b\delta,\gamma+2c\delta,\delta), \quad a+b+c=0, \, a,b,c \in  \mathbb Z$$
and involutions
$$(\alpha,\beta,\gamma,\delta)\rightarrow (-\alpha+2a'\delta,-\beta+2b'\delta,-\gamma+2c'\delta,\delta), \quad a'+b'+c'=1, \, a',b',c' \in  \mathbb Z.$$

There exists another transformation, which swaps $\kappa$ and $t$ and can be considered as a natural candidate for the universal level-rank duality 
\beq
\label{level-rank}
\sigma: (\alpha,\beta,\gamma,\delta)\rightarrow (-\alpha,-\beta,\delta-\gamma,\delta),
\eeq
which is not a symmetry of (\ref{univer}) except for $SU(N)$ case.
To find a universal form of level-rank duality is a very important open question.

\section{Duality of $SO(2n)$ and $Sp(2n)$ Chern-Simons theories}

On the Vogel's plane the coordinates of $SO(2n)$ and $Sp(2n)$ go into each if we interchange $\alpha$ and $\beta$ and change $n$ to $-n.$ This means that on the level of the universal quantities we have formal equality $$Sp(2n)=SO(-2n),$$ which was known for quite a while (see \cite{Ki, Mkr, Cvitbook, VM}). Let's discuss this in a more detail.

Recall that the integer values of the coupling constant $\kappa=k \in \mathbb Z$ correspond to the minimal bilinear form $Tr$ in (\ref{CSA}).
One can check that if one would like to use the trace $Tr_F$ in the defining fundamental representation instead of minimal form with the same quantisation condition, then the corresponding actions will be:
$$SU(N): S(A)=\frac{k}{4 \pi} \int_M  Tr_F \left( A \wedge d A + \frac{2}{3} A\wedge A \wedge A \right)
$$ 
$$SO(N): S(A)=\frac{k}{8 \pi} \int_M  Tr_F \left( A \wedge d A + \frac{2}{3} A\wedge A \wedge A \right)
$$ 
$$Sp(2n): S(A)=\frac{k}{4 \pi} \int_M  Tr_F \left( A \wedge d A + \frac{2}{3} A\wedge A \wedge A \right)
$$ 
Note the additional factor $1/2$ in the $SO(N)$ case. 

According to \cite{Mkr} Yang-Mills gauge theories with 
$$S(A)= -\frac {1}{g^2} \int Tr_F(F_{\mu \nu}F^{\mu \nu})$$
for $SO(2n)$ and $Sp(2n)$ gauge groups are connected by duality transformation $n\rightarrow-n, g^2\rightarrow-g^2.$  The proof works for the Chern-Simons theory as well. 
Now we  have to take into account that in Chern-Simons case one has a distinguished normalisation of the coupling constant, for which $\kappa=k$ is integer, and that normalisation differ by 2 for SO and Sp cases, as shown above. So, we see that  the old duality \cite{Mkr} of  $SO(2n)$ and $Sp(2n)$ gauge theories under  $n\rightarrow-n, g^2\rightarrow-g^2$ becomes $$n\rightarrow-n, \,\, k\rightarrow-2k$$ for the corresponding Chern-Simons theories. Unitary case $SU(N)$ is self-dual under $N \rightarrow -N, k \rightarrow -k.$

This is in perfect agreement with the expressions for the central charge (\ref{cg1}) for $SO(2n)$ and $Sp(2n)$:
$$
SO(2n): \bigskip c=k \frac{ n(2n-1)}{k+2n-2},
$$
$$
Sp(2n): \bigskip c=k \frac{ n(2n+1)}{k+n+1}
$$
are related by $
n \rightarrow -n, k \rightarrow -2k,$
as it should be since the central charge is a universal quantity.

The apparent non-duality of partition functions for $SO$ and $Sp$ theories calculated in \cite{SV} is the consequence of their choice of the invariant forms, where the factor $1/2$ was lost in $Sp$ case. One can show that the same calculation with correct form reveals the duality. 

\section{Concluding remarks}

We discuss some natural questions here. 
The main one is which sectors of the Chern-Simons theory are universal. In particular, is the non-perturbative part of the partition function universal? What's about knotted Wilson loops in representations different from adjoint? 

One can ask also about universality of Yang-Mills theory. Universality of perturbative Chern-Simons partition function as well as universality of Casimir eigenvalues are indications of the universality of perturbative partition function of Yang-Mills theory.

In a universal gauge theory one can consider an analogue of the $1/N$ expansion in two parameters, say $\alpha/\gamma, \beta/\gamma$. In particular, one can consider such expansion for the exceptional Lie groups living on the line (plane) $\gamma=2\alpha+2\beta$ studied in detail by Deligne et al \cite{Del, DM}. 
One can ask also for a universal form of the duality between topological strings and Chern-Simons theory with classical groups \cite{SV, Pe, GV}.

As we have mentioned in the Introduction our results can be interpreted within the theory of affine nontwisted algebras \cite{Kac}. For example, $Z(S^3)=S_{0,0}$ and  $<W(C)>=S_{0,\theta}/S_{0,0}$, where $S_{0,0}$ and $S_{0,\theta}$ are matrix elements of modular transformations of affine characters (see e.g. \cite{DiF}). Parallel will become complete if we include the value of the central element in the definition of affine Lie algebra. Can one extend this picture to the twisted affine Lie algebras? What is the correct form (if any) of the universal level-rank duality?

Note that Lie superalgebras were in Vogel's map from the very beginning \cite{V0,V}, see Table 2.
The meaning of our calculations in that case is still to be understood.

 \begin{table}[h]  
\caption{Vogel's parameters for basic classical Lie superalgebras}     
\begin{tabular}{|c|c|c|c|c|}
\hline
 Lie superalgebra  & $\alpha$ & $\beta$ & $\gamma$  & $t$\\   
\hline    
$\mathfrak {sl}_{m,n}$     & $-2$ & 2 & $m-n $ & $m-n$\\
$\mathfrak {osp}_{p,q}$    & $-2$ & 4& $p-q-4 $ & $p-q-2$\\
$\mathfrak {f}_{4}$    & $-2$ & 2& $3$ & $3$\\
$\mathfrak {g}_{3}$    & $-2$ & 2& $2$ & $2$\\
$\mathfrak {D}_{2,1,\lambda}$    & $\lambda_1$ & $\lambda_2$& $\lambda_3$ & $0$\\
\hline  
\end{tabular}
\end{table}

What are other interesting points on Vogel's map? The theory of Vassiliev knot invariants \cite{CDM}, in particular, the results of \cite{Lieb,HV} may help to understand this.

\section{Acknowledgements.}

RM is grateful to the London Mathematical Society for the support of his visit to the UK in February 2012, when this work was completed, and to A. Tseitlin and H. Khudaverdian for the hospitality. AV is grateful to A.V. Bolsinov, G. Felder and A.N. Sergeev for helpful discussions and to Hausdorff Research Institute for Mathematics, Bonn for the hospitality in January-March 2012.

We would like to thank also H. Mkrtchyan, whose computer program for the root systems calculation was used.


\begin{thebibliography}{99}

\bibitem{V0}
P.Vogel {\it Algebraic structures on modules of diagrams.} Preprint (1995). 
J. Pure Appl. Algebra {\bf 215} (2011), no. 6, 1292-1339. 

\bibitem{V}
 P. Vogel {\it The universal Lie algebra.} Preprint (1999).

\bibitem{H1}
G. 't Hooft {\it A planar diagram theory for strong interactions.} Nucl.Phys. {\bf B72} (1974), 461-473.

\bibitem{W2}
E. Witten {\it Analytic continuation of Chern-Simons theory.} arXiv:1001.2933 (2010).

\bibitem{Ki}
R. C. King {\it The dimensions of irreducible tensor representations of the orthogonal and symplectic groups.} Can. J. Math. {\bf 23} (1971), 176.  

\bibitem{Mkr}
R.L. Mkrtchyan {\it The equivalence of $Sp(2N)$ and $SO(-2N)$ gauge theories.} Phys. Lett. {\bf 105B} (1981), 174-176.

\bibitem{Cvitbook}
P. Cvitanovic {\it Group Theory.} Princeton Univ. Press, Princeton, NJ, 2004. 
http://www.nbi.dk/group theory

 \bibitem{VM}
R.L. Mkrtchyan, A.P. Veselov {\it On duality and negative dimensions in the theory of Lie groups and symmetric spaces.} J. Math. Phys. {\bf 52} (2011), 083514

\bibitem{Ho}
M. El Houari {\it Tensor invariants associated with classical Lie algebras: a new classification of simple Lie algebras.} Algebras Groups Geom. {\bf 14} (4) (1997), 423-446.


\bibitem{An}
E. Angelopoulos {\it Classification of simple Lie algebras.} Panamer. Math. J. {\bf 11} (2) (2001) 65-79.

\bibitem{Del}
P. Deligne {\it La s\'erie exceptionnelle des groupes de Lie.} C. R. Acad. Sci. Paris, S\'erie I {\bf 322} (1996), 321-326.

\bibitem {DM}
P. Deligne and R. de Man {\it La s\'erie exceptionnelle des groupes de Lie II.} C. R. Acad. Sci. Paris, S\'erie I  {\bf 323} (1996), 577-582. 


\bibitem{W1} 
E. Witten {\it Quantum field theory and the Jones polynomial.} Comm. Math. Phys. {\bf 121} (1989),  351-399.

\bibitem{BarNatan1}
D. Bar-Natan {\it Perturbative Chern-Simons Theory.} Journal of Knot Theory and its Ramifications, 4-4 (1995), 503Ð548.

\bibitem{BarNatan2}
D. Bar-Natan {\it On the Vassiliev knot invariants.} Topology, {\bf 34} (1995), 423-472.

\bibitem{RT}
N. Reshetikhin, V.G. Turaev {\it  Invariants of 3-manifolds via link polynomials and quantum groups.} Invent. Math. {\bf 103} (1991), no. 3, 547Ð597.

\bibitem{T}
V.G. Turaev {\it Quantum invariants of knots and 3-manifolds.} Walter de Gruyter and Co., Berlin, 2010. 

\bibitem{LM1} 
J.M. Landsberg, L. Manivel {\it A universal dimension formula for complex simple Lie algebras.} Adv. Math. {\bf 201} (2006), 379-407

\bibitem{LM2}
J. M. Landsberg, L. Manivel {\it Triality, exceptional Lie algebras and Deligne dimension formulas.} Adv. Math. {\bf 171} (2002), 59-85.

\bibitem{MSV}
R.L. Mkrtchyan, A.N. Sergeev, A.P. Veselov {\it Casimir eigenvalues for universal Lie algebra.} arXiv:1105.0115 (2011).

\bibitem{FdV}
H. Freudenthal and H. de Vries {\it Linear Lie Groups.} Academic Press, 1969.


\bibitem{DiF} 
P. Di Francesco, P. Mathieu, and D. Senechal {\it Conformal Field Theory.} Springer-Verlag, New York, 1997. 

\bibitem{OV}
H. Ooguri and C. Vafa {\it Worldsheet derivation of a large N duality.} Nucl. Phys. {\bf B 641} (2003), hep-th/0205297.

\bibitem{Kac} 
V. Kac {\it Infinite dimensional Lie algebras.}Third Edition, Cambridge University Press, 1995.



\bibitem{Ma}
I. G. Macdonald {\it The volume of a compact Lie group.} Invent. Math. {\bf 56}(1980), 93-95.

\bibitem{Mar}
M.S. Marinov {\it Invariant volumes of compact groups.} J. Phys. A: Math. Gen. {\bf 13} (1980), 3357-3366.

\bibitem{Hash}
Y. Hashimoto {\it On Macdonald's formula for the volume of a compact Lie group.}
Comment. Math. Helv. {\bf 72} (1997), 660-662.

\bibitem{Bour}
N. Bourbaki {\it Groupes et Alg\`ebres de Lie.} Chap.~VI, Masson, 1981.

\bibitem{H}
G. Harder {\it A Gauss-Bonnet formula for discrete arithmetically defined groups.} Ann. sci. de l'\'Ecole Normale Sup\'erieure, S\'er. 4, 4 no. 3 (1971), p. 409-455 

\bibitem{SV}
S. Sinha and C. Vafa {\it SO and Sp Chern-Simons at Large N.} hep-th/0012136 (2000).

\bibitem{NS}
S.G. Naculich, H.J. Schnitzer {\it Level-rank duality of the U(N) WZW model, Chern-Simons theory, and 2d qYM theory.} J. High Energy Phys. (2007), no. 6, 023, 19 pp.
 
 \bibitem{Pe} 
V. Periwal {\it Topological closed string interpretation of Chern-Simons theory.} Phys. Rev. Lett. {\bf 71} (1993), 1295; hep-th/9305115.

\bibitem{GV} R. Gopakumar, C. Vafa {\it On the Gauge Theory/Geometry Correspondence.} Adv.Theor.Math.Phys. {\bf 3} (1999) 1415-1443, arXiv:hep-th/9811131v1.

\bibitem{CDM}
S. Chmutov, S. Duzhin and J. Mostovoy {\it Introduction to Vassiliev knot invariants.} arXiv:1103.5628, 
Cambridge Univ. Press, Cambridge, 2011.    

\bibitem{Lieb}
J. Lieberum {\it On Vassiliev invariants not coming from semisimple Lie algebras.} J. Knot Theory Ramifications 8 (5) (1999) 659–666. 

\bibitem{HV}
V. Hinich, A. Vaintrob {\it Cyclic operads and algebra of chord diagrams.} Selecta Math. (N.S.) 8 (2002), no. 2, 237–282.

\end{thebibliography}
\end{document}